# Atomically Resolved Imaging of Highly Ordered Alternating Fluorinated Graphene

Reza J. Kashtiban[1], M. Adam Dyson[1], Rahul R. Nair[2], Recep Zan[2,3], Swee L. Wong[2], Quentin Ramasse[4], Andre K. Geim[2], Ursel Bangert[5,6], Jeremy Sloan[1]

[1]Department of Physics, University of Warwick, Coventry, CV4 7AL, UK
[2]School of Physics and Astronomy, University of Manchester, Manchester, M13 9PL, UK
[3]Department of Physics, Faculty of Arts and Sciences, Niğde University, Niğde 51000, Turkey
[4]SuperSTEM Laboratory, STFC Daresbury Campus, Daresbury WA4 4AD, UK
[5]School of Materials, University of Manchester, Manchester, M1 7HS, UK
[6]Department of Physics & Energy, University of Limerick, Castletroy, Limerick, Ireland

**One of the most desirable goals of graphene research is to produce ordered 2D chemical derivatives of suitable quality for monolayer device fabrication. Here we reveal, by focal series exit wave reconstruction, that $C_2F$ chair is a stable graphene derivative and demonstrates pristine long-range order limited only by the size of a functionalized domain. Focal series of images of graphene and $C_2F$ chair formed by reaction with $XeF_2$ were obtained at 80 kV in an aberration-corrected transmission electron microscope. EWR images reveal that single carbon atoms and carbon-fluorine pairs in $C_2F$ chair alternate strictly over domain sizes of at least 150 nm$^2$ with electron diffraction indicating ordered domains $\geq 0.16 \mu m^2$. Our results also indicate that, within an ordered domain, functionalization occurs on one side only as theory predicts. Additionally we show that electron diffraction provides a quick and easy method for distinguishing between graphene, $C_2F$ chair and fully fluorinated stoichiometric CF 2D phases.**

The extraordinary structural and transport properties of graphene[1] have given rise to an intense interest in its morphological and chemical modification resulting in an extensive range of derivative materials. This has been driven by the consideration that graphene itself must be modified to achieve a usable band gap and other desirable low-dimensional properties. One approach is the nano-engineering of graphene to form nano-ribbons so that charge carriers are confined to a quantum wire[2,3]. A more scalable approach is the formation of chemical derivatives such as graphene oxide (GO)[4,5], hydrogenated graphane (CH)[6] or fluorinated graphene ($C_xF$, $x \leq 4$)[7-9]. GO consists of graphene sheets decorated with epoxy, hydroxyl and carboxyl groups[4] whereas graphane is hydrogenated graphene[6]. For some derivatives, structural order on length scales further than a few C-C bond distances cannot be demonstrated and for GO, even the local structure remains a matter for significant debate[4]. For graphane, no long-range order is observed due to the absence of uniformity in the corrugations of the benzene ring[7] a problem compounded by the low stability of this structure at moderate temperatures[6].

Stoichiometric fluorographene[8], a graphenic monolayer with each fluorine atom bonded to a carbon atom in a distorted $sp^3$ sheet, would appear to be the most likely candidate for a usable graphene derivative. This material is a thermally and chemically stable insulator with similar mechanical strength to graphene offering a range of possible applications[8-13]. However the reported 2D lattice constant for CF is ~0.248 nm which is apparently expanded only 1% relative to graphene, significantly lower than the 2.8% expanded lattice constant for monolayer CF predicted by density functional theory (DFT) and notably also less than the 2.8-4.5 % expanded lattice constant variously reported for graphite fluoride[8,10,14,15]. The observed lower lattice parameter reported for CF indicates that this phase may undergo significant lattice corrugation that will impair its utility in any application requiring a "flat" 2D morphology. Here we show, using electron diffraction and aberration corrected-transmission electron microscopy (AC-TEM) *in tandem* with exit wave reconstruction (EWR)[16-19], that the DFT predicted phase $C_2F$ chair[10] is both stable and demonstrates a far higher degree of pristine long-range structural and morphological order than CF or any other chemical derivative of graphene. Our observations also support theoretical predictions[10] that, due to energetic considerations, ordered domains of $C_2F$ chair are functionalized exclusively on one side, a result with profound implications for the preparation of 2D devices and, furthermore, the formation of secondary chemical derivatives such as those formed by alkylation, hydroxylation and amino-functionalization.[20] In addition, new carefully calibrated electron diffraction studies performed on a freshly fluorinated graphene monolayer sample provide a domain-by-domain 2D phase analysis that both supports the conclusion of our EWR studies but also reveals that some domains of fully fluorinated graphene may possibly be uncorrugated.

AC-TEM can provide information about atomic arrangements within materials at low accelerating voltages, reducing specimen damage although images are often noisy and difficult to interpret. High Angle Annular Dark Field (HAADF) imaging performed in Scanning Transmission Electron Microscopy (STEM) produces higher definition images but may degrade thin monolayers due to the high electron flux of the highly focused electron beam. Exit Wave Reconstruction (EWR)[16-19] can recover more information from AC-TEM by combining data from a focal series of low beam density images, providing light element sensitivity and even 3D information[21]. We present here comparative EWR phase images calculated from focal series obtained from both pristine graphene and also partially fluorinated 2D monolayer $C_xF$ ($x$ = 1,2) samples under nearly identical imaging conditions suggesti. This imaging technique permits direct imaging of individual C atoms and alternating C-F atom pairs at atomic resolution and the obtained images also indicate pristine long-range order in this structure. The preservation of the microstructure of $C_2F$ during extended focal series acquisition is furthermore an important indicator of its comparative stability for unperforated $C_2F$ chair monolayers. However we also reveal how prolonged electron beam irradiation of this material in order to form perforated monolayers leads to sequential sputtering of F atoms and C atoms from the terminal edge of a hole leading to a modified decomposition sequence relative to similar sputtering reported for graphene.

**Preparation and general characterisation of monolayer C$_x$F**

CVD-grown graphene (Fig. 1a, Supplementary Fig. 1a) was suspended on TEM 'Quantifoil' specimen grids by applying previously published protocols for the synthesis, transfer and cleaning of this material[22,23]. Monolayer C$_x$F (with possible microstructures indicated in Fig. 1a and Supplementary Figs. 1b-e[10]) was then produced by partially fluorinating additional suspended graphene samples using the previously reported XeF$_2$ direct fluorination method[8,23] (Methods section). While this technique can also be used to produce stoichiometric CF (Fig. 1a, Supplementary Fig. 1d), careful regulation of the extent of fluorination enables partially fluorinated samples to be prepared by adjusting the applied temperature between the range 70 to 200° C[8]. We performed carefully calibrated electron diffraction studies on an initial sample (Fig. 1b and c) and subsequently a freshly prepared fluorinated graphene samples (Supplementary Figs. 2a-d) to test the distribution of 2D phases in both preparations as predicted by DFT by Şahin[10] (Supplementary Tables 1 and 2, ED simulations in Fig. 1d-f and Supplemetary Fig. 1F) as EELS studies performed on the initial sample (Supplementary Figs. 3a) indicated some possible sample deterioration. In the new study, energy dispersive X-ray microanalysis (EDX) was performed simultaneously with the ED study (Supplementary Fig. 3b) and clearly revealed the presence of fluorine in specimens with an enlarged lattice parameter although some contaminaiton was also indicated. Additional Raman studies (Supplementary Fig. 4) suggest that the extent of fluorination is quite variable across this sample with an uneven distribution of sp$^3$ versus sp$^2$ functionalisation although the poor spatial resolution of this method (typically 1 µm$^2$) is unlikely to give a clear picture of the ordering within C$_x$F on a domain-by-domain basis.

**Electron diffraction studies of graphene and C$_x$F**

Electron diffraction (ED) was carried out on graphene and the initial monolayer C$_x$F sample using a diffraction aperture of ~0.16 µm$^2$ and equivalent exposure conditions for both materials. Overlaid ED patterns for pristine graphene (Fig. 1b, white spots) and the C$_x$F phase (Fig. 1b, red spots) both show 6-fold symmetry with the sharp intensities of the latter pattern indicating a high degree of order[24]. Additionally, the ratios of $\bar{2}110_{C2F}$ to $\bar{1}010_{C2F}$-type reflections are consistent with those obtained for graphenic monolayers[24], whereas the measured unit cell revealed a 2.4% expanded lattice parameter (i.e. $a_1, a_2 = 0.251 \pm 0.008$ nm), larger than the 0.248 nm parameter reported for 'corrugated' CF[8] although less than the value predicted for 'uncorrugated' CF (Supplementary Table 1).[10] The indicated unit cell is also in good agreement with the unit cell predicted by DFT for $C_2F$ 'chair' functionalised on one side only (i.e. $a_1, a_2 = 0.252$ nm).[10] The relative intensities of $10\bar{1}0$-type reflections for graphene and the $C_2F$ phase scale in a manner consistent with adding an additional F atom per $C_2F$ unit cell relative to graphene (Fig. 1c)[25]. No domains with electron diffraction corresponding to the tetragonal 'boat' form of $C_2F$, also predicted by DFT (with lattice parameters $a_1 = 0.254$ nm; $b_1 = 0.436$ nm[10]), were observed which can readily be distinguished from the chair form of $C_2F$ by ED simulations generated from both DFT predicted forms (i.e. *cf.* Figs. 1b,e and f). ED patterns obtained from about 20 regions of the initial $C_xF$ sample indicated that ~70% of the observed domains had a similar microstructure with EWR studies indicating that many of these exhibited short to long range order. Stoichiometric CF has not so far been identified in this initial $C_xF$ sample.

Following careful calibration of the camera length of our electron microscope with a polycrystalline gold sample (Supplementary Fig. 2a) we were able to distingush between different domains (or 2D phases) of graphene, $C_2F$ chair and could even identify stoichiometric CF (with the enlarged 0.255 nm lattice parameter as predicted by Şahin[10]) for 18 different fragments for a freshly prepared $C_xF$ sample with a high degree of confidence. This sample was found to contain domains of unfluorinated graphene and 2D phases with a lattice parameter conforming to $C_2F$ chair as described above. The distribution of 2D phases in this sample was found to be 7:9:2 for graphene, $C_2F$ chair and stoichiometric CF respectively (Supplementary Tables 1 and 2). These results suggest that one-sided fluorination is rapid and completely 'coats' one side of graphene with fluorine before the opposing side is fluorinated, a process which may be initiated either catalytically or by defects.

**Imaging of graphene and $C_2F$ chair by Exit Wave Reconstruction**

Exit Wave Reconstruction (EWR) was then employed on domains of graphene and $C_2F$ chair exhibiting comparable ED behaviour to Figs. 1d and e and also in terms of the enlarged ~0.252 nm lattice parameter. For both structures, focal series consisting of 34 images were obtained at 80 kV in an AC-TEM (Methods section, Supplementary Figs. 5 and 6). The image in Fig. 2a shows the restored phase for a ~100 $nm^2$ graphene domain obtained with ~0.09 nm resolution and details obtained from the EWR and false-colour images (Figs. 2b and c) produced by thresholding the phase in order to reveal the positions of the C atoms (highlighted in blue) comprising the graphene lattice. Line profiles obtained along C-C dumbells (see also Figs. 3a and c) reveal bonds of ~0.142±0.008 nm, consistent with monolayer graphene. In Figs. 2d,e and f, an EWR phase image, detail and false-colour image obtained from a ~144 $nm^2$ domain of $C_2F$ is then shown. The important outcome of the EWR images is the heightening of the phase shift due to –CF< pairs relative to the >C- atoms as suggested by the inset model in Fig. 2e. As above, the false-color map in Fig. 2f indicates the positions of the single C atom columns in blue while the >CF- pairs indicated by green peaks occur on strictly alternating C positions. In Figs. 2g, h and i we show low magnification, detail and thresholded simulated EWR images of domains of the fluorographene equivalent in size to the experimental images reproduced in Figs 2d-f are in good agreement with these images.

In Figs. 3a-f a more detailed interpretation of enlarged 8×17 nm domains of both experimental and simulated phase images is shown (Figs. 2d and g). The two models in Figs. 3a and b correspond to domains of graphene and $C_2F$ chair as in Figs. 3c and d. Shown side-by-side with the experimental images are simulations computed from the models in Figs 3a and b. Figs. 3e and f show overlaid line profiles corresponding to the diffuse lines in Figs. 3c and d. In Fig. 3e, the phase image and simulations produce clearly defined dumb-bells for monolayer graphene with peaks separated by 0.142 nm, equivalent to the graphenic >C–C< separations. Experimental and simulated line profiles for $C_2F$ in Fig. 3f for three >C–CF< dumb-bells present with saw-tooth profiles with the low peaks corresponding to single >C- atoms and the taller peaks to >C–F pairs.

These are separated by ~0.146 nm, equivalent to the tetrahedrally distorted 0.148 nm >C–CF< distance (predicted by DFT[10]) imaged in a 'plan view' projection (*cf.* $C_2F$ chair in Fig. 1a)[10]. Computing the experimental >C–CF< distance from the reported buckling (δ) of 0.029 nm for $C_2F$ chair from the average experimental 'plan view' distance gives ~0.149 nm, consistent with the reported DFT value for the >C–CF< distance within experimental error. In addition, the relative magnitude of the experimentally determined and simulated atom column phase shifts for single –C< columns and >C-F pairs give a clear distinction between carbonaceous graphene and $C_2F$ chair and excellent agreement with simulated phase shifts computed for the aberration values determined for our instrument (i.e. Figs. 3g and h, see also Methods section). The relative phase shifts and spatial distributions of these features also readily allow us to distinguish between this microstructure and that for the boat form of $C_2F$ as well as that for stoichiometric fluorographene by performing comparisons with simulations obtained for all four structural forms taking into account net phase shifts for single C columns versus corresponding shifts for >C-F pairs (see Supplementary Figs. 7 and 8).

The line profiles in Fig. 3f reveal that the functionalization of $C_2F$ chair exhibits outstanding short-range order. Wider field of view images in Fig. 2d, Figs. 4a,b and also Figs. 5a,b give indications of how this order is retained for larger sized domains. In the false-colour surface profile image in Fig. 4c (generated from Fig. 4b) the peaks corresponds to the measured phase shift in C and CF atom columns for a ~64 $nm^2$ area of $C_2F$ chair. A triangulation of three 8 pixel line profiles (**I**, **II** and **III**, Fig. 4b) through a total of 51 >CF–C< or >C–CF< dumb-bells is reproduced in Fig. 4d. The taller –CF< peaks predominate in the profile image and over the entire domain the microstructure maintains pristine order. A comparison of this region with the wider field of view image (Fig. 4a) indicates that the ordering extends well beyond the range of this image possibly to regions equivalent in size to the ED aperture (~0.16 $\mu m^2$). Notably there is little rippling in the domains in Figs. 2 and 4 although small ripples are present in the wider field of view image indicating that the significant corrugation reported for stoichiometric $CF^8$ is absent. A larger ~150 $nm^2$ ordered domain image (Supplementary Fig. 9) also presents with little rippling and, most significantly, with no evidence of short or long-range disorder or strain effects associated with local disruption of the $C_2F$ chair 2D lattice.

**Prolonged in situ electron beam irradiation studies of $CF_x$**

We have also investigated the stability of the $C_2F$ chair microstructure following prolonged exposure to an electron beam at 80 kV (Fig. 5a-g). During sample irradiation, a region of monolayer $C_2F$ chair material was exposed to an electron beam density of ca. $10^6$ e$^-$/nm$^2$ for 20 minutes after which both focal series and 'single shot' AC-TEM images (Fig. 5d) were recorded from regions of hole formation (see Methods section). In Figs. 5a-c an EWR phase image and details obtained from a typical hole are shown. A few C-C bond distances in from the edge of the hole, the microstructure of the $C_2F$ chair monolayer is perfectly retained while at the periphery the contrast is somewhat reduced, possibly indicating preferential removal of single fluorine atoms from the edges (which can be induced more systematically[26]). In the edge enlargement in Fig. 5c, we see that there are some residual carbon fragments (arrowed) but both focal series images and image series obtained at optimum defocus (Fig. 5d) indicate dynamical rearrangement of the microstructure at the edges (Fig. 5e). The first 1-3s exposures following prolonged illumination reveals the presence of reconstructed zig-zag ('reczag') features consisting of 5- and 7-membered ring pairs which has been reported for holes formed from pristine graphene although migration of C-C atoms pairs along graphene edges is more common.[27,28] We also note that these edge features are mobile, and a partial shunt of all the reczag units (resulting in the formation of a single Klein edge feature[27]) along the edge is observed (i.e. Fig. 5f) as suggested by the models in Fig. 5g just prior to elimination of the whole row from the edge of the hole after 5s. Taken together, these results underscore the comparative stability of the $C_2F$ chair monolayers that only degrade following prolonged electron beam exposure at 80 kV. Elimination of F from the $C_2F$ chair structure at the hole edges results in more graphene-like behaviour and similar edge reconstruction behavior to that previous reported for graphene in particular.

**Discussion**

ED and EWR both reveal that $C_2F$ chair is an ordered 2D monolayer material but how stable is this structure and how reliable is our assertion of one-side functionalization ? The robustness of the $C_2F$ chair 2D phase in the electron beam during focal series acquisition is a strong indicator of its stability, especially in comparison to other part-functionalized graphene materials. Erni et al.[29] demonstrated that ad-atoms and ad-molecules chemically attached to graphene can be imaged by EWR in an AC-TEM using focal series of up to 11 images with comparable exposure conditions to those employed here (see Methods section). We were able to take series of up to 34 images for EWR images for domain sizes >64 nm$^2$ (i.e. Figs. 4a-c) with no evidence of significant rearrangement of the local microstructure although eventually the $C_2F$ chair monolayers do degrade but only following prolonged beam exposure at 80 kV for 20 minutes or more. This latter behavior is wholly consistent with similar observations reported for both pristine and other chemically modified graphenic monolayers[27,28].

The assertion that fluorine functionalization of $C_2F$ chair occurs on one side is justified by the bond lengths derived from ED patterns and reported theory[10] even though both sides of the graphene were simultaneously exposed to $XeF_2$[8]. In addition, our electron diffraction study on a freshly fluorinated graphene sample reveals that unfluorinated graphene domains, partially fluorinated $C_2F$ chair and fully fluorinated graphene (designated **Gr**, ***Ch.*-$C_2F$ *Stoich*-CF**, in the accompanying schematic in Supplementary Table 2). In their DFT study, Şahin *et al.* indicate that $C_2F$ chair functionalized on one side has the lowest formation energy relative to graphene (i.e. 0.09 eV) in comparison to 1.44 eV for $C_4F$, 0.91 eV for $C_2F$ boat and 2.04 eV for CF[9]. While $C_2F$ boat may be inherently more stable, formation of this structure requires fluorination on adjacent C atoms whereas $C_2F$ chair requires fluorination on alternating C atoms, a less sterically hindered process (*cf.* $C_2F$ chair and boat models, Fig. 1a). Additionally once fluorination nucleates on one side, it may progress energetically downhill until the domain functionalized on one side before obverse fluorination occurs. If true, our work provides support for the proposed one-sided stepwise fluorination sequence suggested for stoichiometric CF[10] and the observation of well-defined graphene, $C_2F$ chair and stoichiometrically fluorinated domains of CF provide strong support for this. We also note that a triangular lattice with F sitting on top *and* below would be a frustrated system, a classical example of which is spin ice[30] and, additionally, local deviation from one-sided functionalization would result in readily observable topological distortions in the $(C_2F)_3$ chair rings and longer-range strain effects.[31] In none of the experimental images do we see such distortions and the $(C_2F)_3$ chair rings all retain an undistorted hexagonal shape. Additionally, the lack of disorder in the ED patterns (Fig.1b) is further evidence for the lack of distortion or other forms of disorder.

We note that other modes of fluorination have been reported. Robinson *et al.* reported one-sided fluorination at 1,4 positions on the graphene rings to form $C_4F$ when graphene was initially exposed to $XeF_2$ on one side.[32] This material was then converted to stoichiometric CF following subsequent fluorination of the interim product on both sides. Lee *et al.*[33] presented calculations in support of this although these do not address out-of-plane distortions that may be key in terms of establishing the fluorination mechanism. We unequivocally observe only 1,3,5 fluorinated graphene $C_2F$ chair domains in our partially fluorinated samples and we find no evidence for a $C_4F$ ordered superstructure (Supplementary Figure 1e and d. See also Ref. 10) in our partially fluorinated material. If both cases are true, this suggests that different fluorination mechanisms may be achieved by subtle alteration of the reaction conditions as indicated by Şahin *et al*[10]. Additional microstructural studies may help to confirm the alternative fluorination mechanisms suggested by Robinson *et al* .[32]

In conclusion, we have for the first time characterized both the structure and stoichiometry of an alternating fluorinated graphene material with atomic resolution using EWR. $C_2F$ chair is a highly ordered material that demonstrates selective alternating fluorination on one side for domains >150 nm$^2$ in accord with previous theoretical work.[10] Our results indicate that preferential functionalization of graphene by fluorine on one side appears to be energetically favoured even when graphene is exposed on both sides to $XeF_2$[8,32]. The observed single-sided domains are likely to self-organise due to the finite mobility of F-atoms along graphene[10], resulting in clean patches being randomly fluorinated on the top or bottom sides. The strongly electronegative character of the >C–CF< functionals combined with the highly anisotropic nature of mono-sided functionalization indicate a significant potential for creating ordered secondary derivatives from $C_2F$[20,34]. Furthermore, $C_2F$ chair presents with an undistorted 2D morphology in contrast with stoichiometric but corrugated CF with the consequence that the former is a potentially much more tractable material for 2D device fabrication[8].

**Methods**

**Preparation of graphene and $C_2F$ films for AC-TEM** CVD grown graphene 2D crystals used for the comparative imaging study of this material were first synthesized, transferred onto Quantifoil AC-TEM grids and then cleaned according to standard published protocols[22,23]. A second Quantifoil-suspended CVD graphene sample was also prepared for fluorination using the same method. Fluorination was performed on this second sample by direct fluorination with $X_2F$ gas in a Teflon container at 70° C[8]. Raman spectroscopy was performed on as-prepared fluorinated CVD graphene membranes prepared on TEM grids before performing the TEM experiments. These studies were performed using a Renishaw spectrometer equipped with a 514 nm laser and using a ~1 μm diameter spot.

**Exit wave reconstruction and simulation**

A JEM-ARM 200F microscope operating at 80kV equipped with a CEOS aberration corrector and a Gatan SC1000 ORIUS camera with a 4008×2672 pixel CCD was been used for TEM investigations. A Gatan fiber-optical coupled SC1000 ORIUS camera with CCD size of 4008 by 2672 pixel was used for image acquisition. EWR was carried out using 34 image through focal series with focal steps of ~1.5 nm and a sampling rate of 0.00811 nm/pixel, satisfying the Nyquist criterion. Electron beam densities were adjusted in order to be similar to those reported in Ref. 26 (i.e. $10^6$ e⁻/nm$^2$). Typical values for the residual aberrations of the JEOL ARM 200F were recorded as follows Defocus: (C1) = - 318 ± 2 nm; twofold astigmatism: (A1) = 6 ± 2 nm, threefold astigmatism: (A2) = 44 ± 10 nm, coma: (B2) = 22 ± 10 nm, 3rd order spherical aberration: (C3) = 1.22 ±08 μm, fourfold astigmatism: (A3) = 390 ± 100 nm, star aberration: (S3) = 1.2 ± 0.2 μm, fivefold astigmatism: (A4) 140 ± 10 μm. It must be noted that these values are acquired at magnification of 500,000 and drift in real time and may undergo further drift when the lattice images of graphene and $C_2F$ chair were obtained. These considerations notwithstanding, typical resolutions were obtained from FWHM measurements of individual atom columns from EWR reconstruction reveal spatial frequencies of ~0.11-0.12 nm with FFTs obtained from individual lattice images indicating a spatial resolution of ~0.094 nm.

The FTSR package by HREM Research was used to perform EWR[20]. AC-TEM simulations of graphene and $C_2F$ were calculated using a finite-difference multislice simulation routine. The graphene and $C_2F$ models were constructed in Mathematica 8.0 (Wolfram Research, Inc.) with bond lengths and angles adopted from DFT models[10]. Simulated EWR images were calculated using parameters matching the experimentally determined C3 for our instrument using a fast multi-slice algorithm as described in 'Advanced Computing in Electron Microscopy' 2nd Ed. by E. J. Kirkland, Springer, 2010. To investigate the stability of the structure under the electron beam a single monolayer of $C_2F$ chair was exposed to electron beam for ~20 min under the same illumination conditions employed for focal series acquisition which produced a hole in the $C_2F$ chair sheet. After hole formation, a focal series of 30 images with focal step of 1.5 nm and sampling rate of 0.00782 nm/pixel and exposure time of 1s per image was acquired and used to reconstruct the exit wave from the drilled hole.

EELS studies were also performed on partially fluorinated graphene samples at the SuperSTEM Laboratory on a Nion UltraSTEM100 dedicated ultrahigh vacuum scanning transmission electron microscopes equipped with cold field emission gun with a native energy spread of 0.3–0.35 eV and operating at 60 keV. EDX studies were performed on monolayer fluorinated graphene samples in the ARM200F AC-TEM using a ~3nm probe and an Oxford Instruments SDD X-ray microanalysis detector. Electron diffraction patterns were obtained in the same instruments and on the same samples using a ~0.16 μm$^2$ selectred area diffraction aperture using a 20 cm camera length and using 400852672 pixel CCD. The latter was calibrated with a polycrystalline Au sample (a typical pattern is recorded in Fig. S2a) similar to the method described in Ref. 35. The precision of the lattice parameter measurement is at least 0.3% (i.e. limited by the Au calibration) although individual reflections on ED patterns recorded from monolayer graphene and fluorographene samples can be measured with a precision of ~0.15%.

**Acknowledgements** We thank the EPSRC for funding through a studentship for M. A. D. and for a P. D. R. A. Fellowship for R. J. K. and additional support provided by the Warwick Centre for Analytical Science (EP/F034210/1).

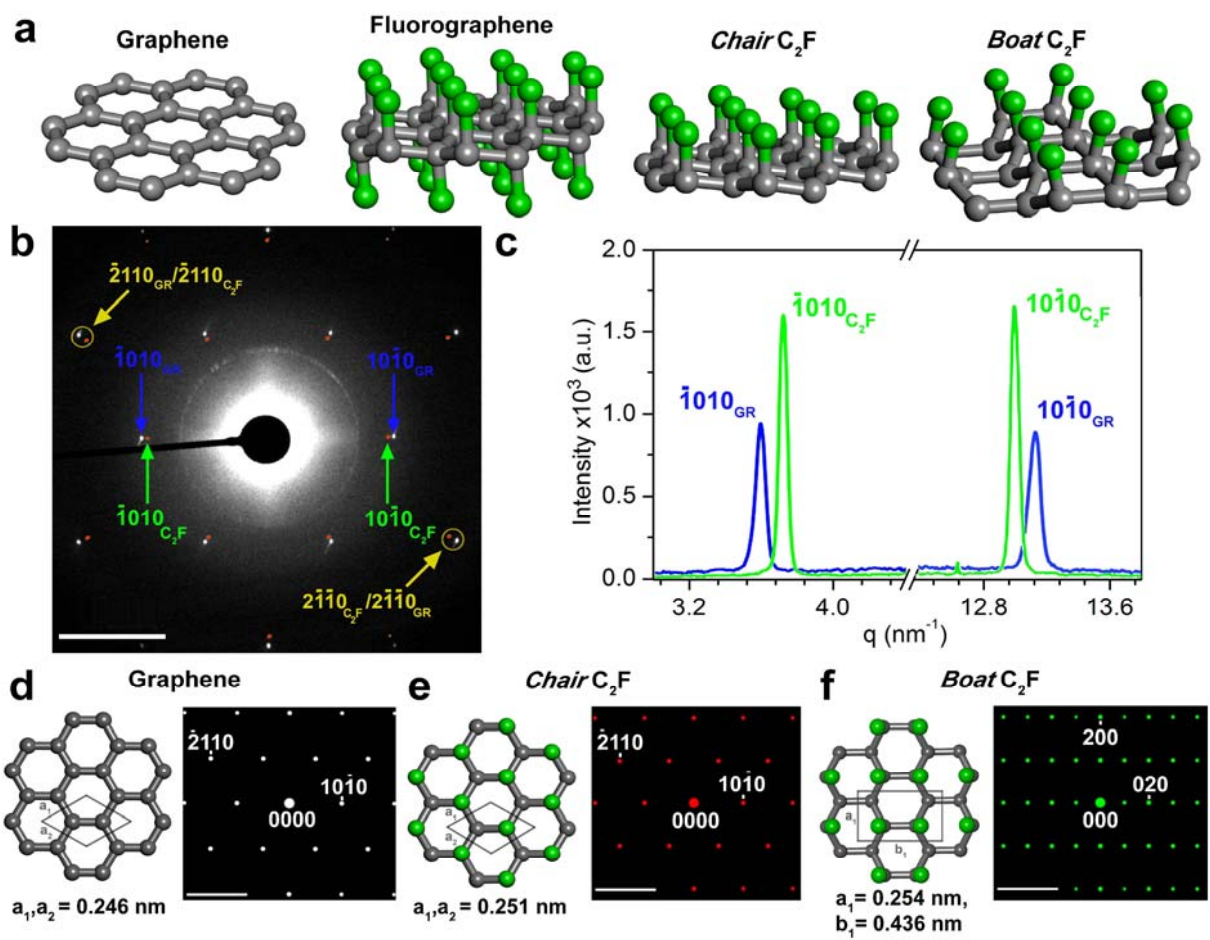

**Figure 1 | Electron diffraction study of pristine graphene and C$_2$F chair. a**, (L-R) Perspective models of graphene, stoichiometric fluorographene, C$_2$F chair and C$_2$F boat. **b**, ED patterns obtained from graphene (subscripted GR) with overlaid scaled ED pattern obtained from monolayer C$_2$F (subscripted C$_2$F) with *hkil* $\bar{1}$010-type and $\bar{2}$110-type reflections indicated for both phases (scale bar = 5 nm$^{-1}$). **c,** Intensity line profiles obtained through $\bar{1}$010 and 10$\bar{1}$0-type reflections for graphene and monolayer C$_2$F. **d,e** Structure model (left), experimentally determined unit cells produced from calibrated ED data in **b** and **c**, and ED patterns (right) for graphene and C$_2$F chair (scale bars = 5 nm$^{-1}$). The estimated precision of the unit cells is ± 0.3%. **f,** DFT determined structure for C$_2$F boat[9] together with simulated ED pattern (*hkl* indices, scale bar = 5 nm$^{-1}$).

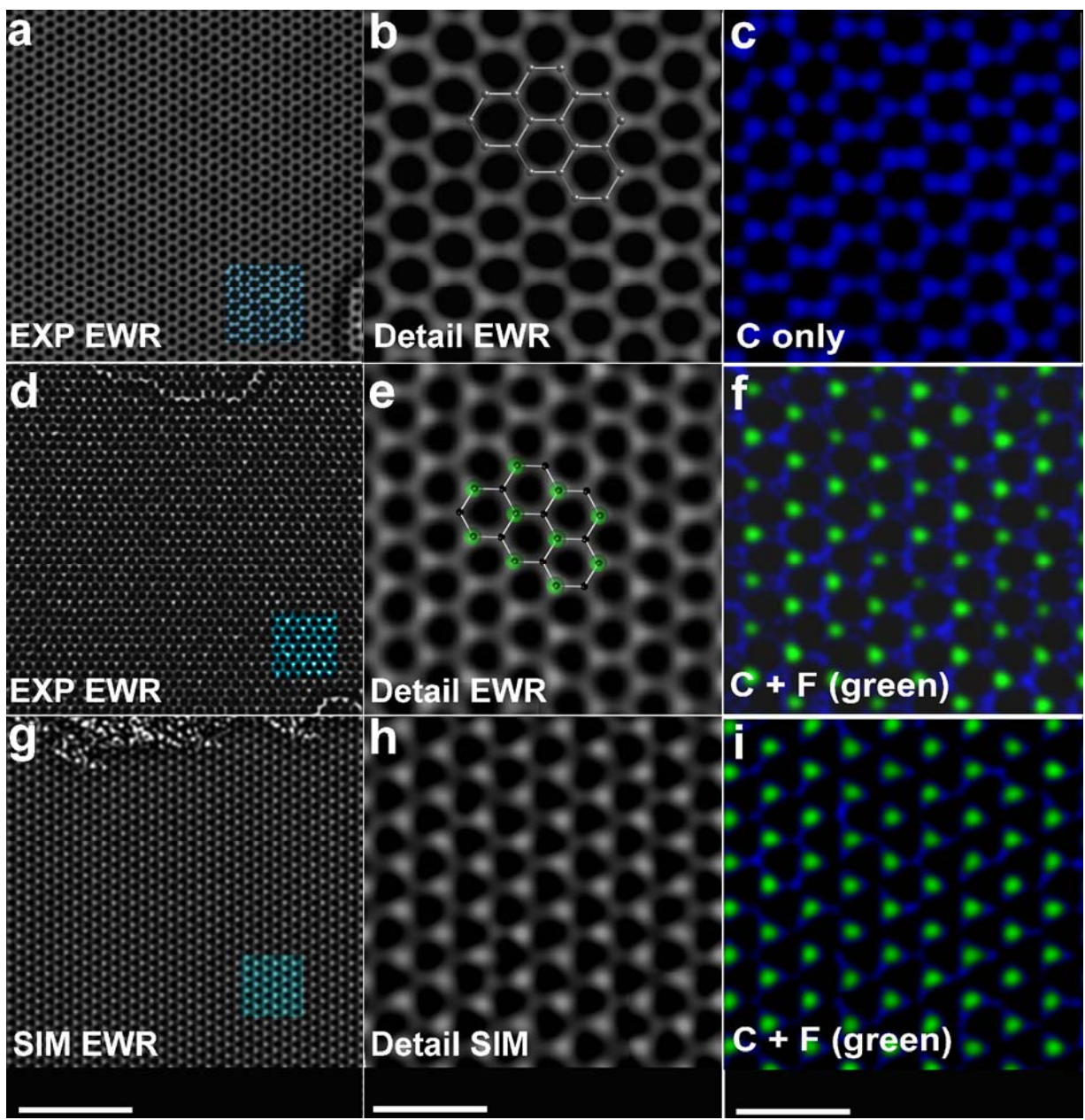

**Figure 2 | Exit wave reconstructions (EWR) and simulated EWRs of pristine graphene and monolayer $C_2F$ chair. a**, Experimental (EXP) restored phase obtained from a ~64 nm$^2$ domain of graphene. **b**, Higher magnification view of the restored phase produced from the highlighted domain in **a** with a graphene 'ball-and-stick' model overlaid. **c,** Thresholding the detail in **b** produces a false-colour plot in which the blue spots corresponds to the phase shift produced by individual C atoms in graphene. **d-f,** as **a-c** but for $C_2F$ with F atoms separately highlighted in green. **g-i,** Simulated phase images for $C_2F$ chair. The intensity distribution of the simulation provides an excellent match with the experimental phase image in **f**. (Parts **a**, **d**, **g** scale bar = 3 nm; Parts **c**, **d**, **e**, **f**, **h**, **i** scale bar = 0.5 nm)

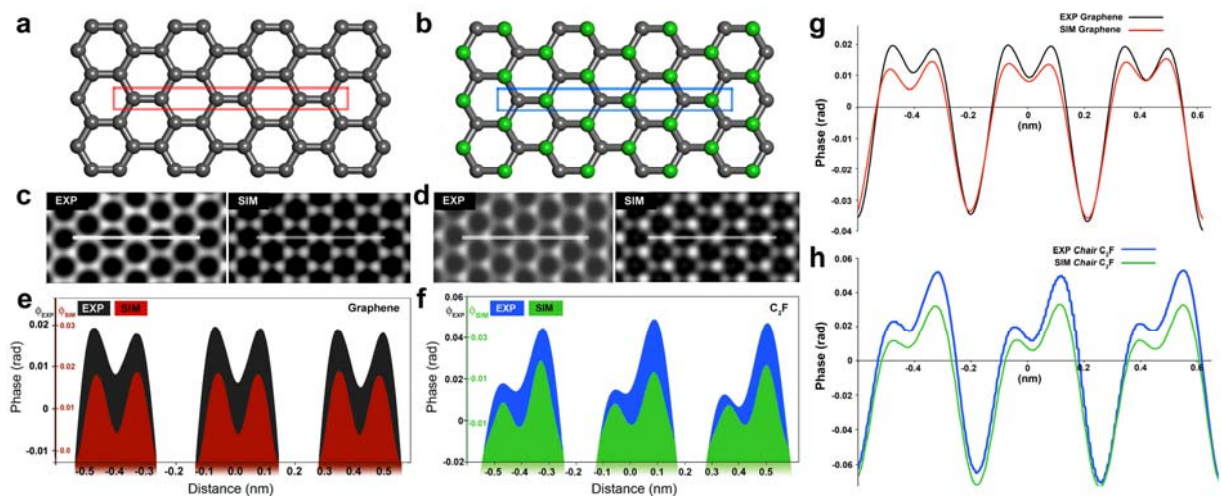

**Figure 3 | Semiquantitative comparisons between experimental and simulated EWR images of pristine graphene and $C_2F$ a-b,** Graphene and $C_2F$ models with C-C and C-CF dumb-bells highlighted. **c-d,** Equivalent domains of experimental (EXP) and simulated (SIM) phase images for monolayer graphene and $C_2F$ chair respectively. **e,f** Colour-coded line profiles obtained from the indicated regions in **c** and **d** for the EXP and SIM phase images for graphene and $C_2F$ chair. The line profiles obtained from the SIM images are artificially downshifted by ~0.02 rad for clarity. The peaks in **e** correspond to three graphene dumb-bells whereas the three saw-tooths in **f** corresponds to >C– atoms (low peaks) and –CF< pairs (tall peaks) in a strict >C–CF< sequence. **g** Overlaid full plots of the experimental (EXP) and simulated (SIM) phase contrast for pristine graphene, respectively. **h** Overlaid full plots of the experimental (EXP) and simulated (SIM) phase contrast for $C_2F$ chair, respectively. Note that the net phase shift for single C atoms in graphene (i.e. $|\phi|$) ~0.5 rad differs from the net phase shift for single C atoms in $C_2F$ chair (~0.08 rad) due to the convolution of this with the net phase shift for –CF< pairs (i.e. ~0.1 rad). See also Supplementary Figs. 8a and b.

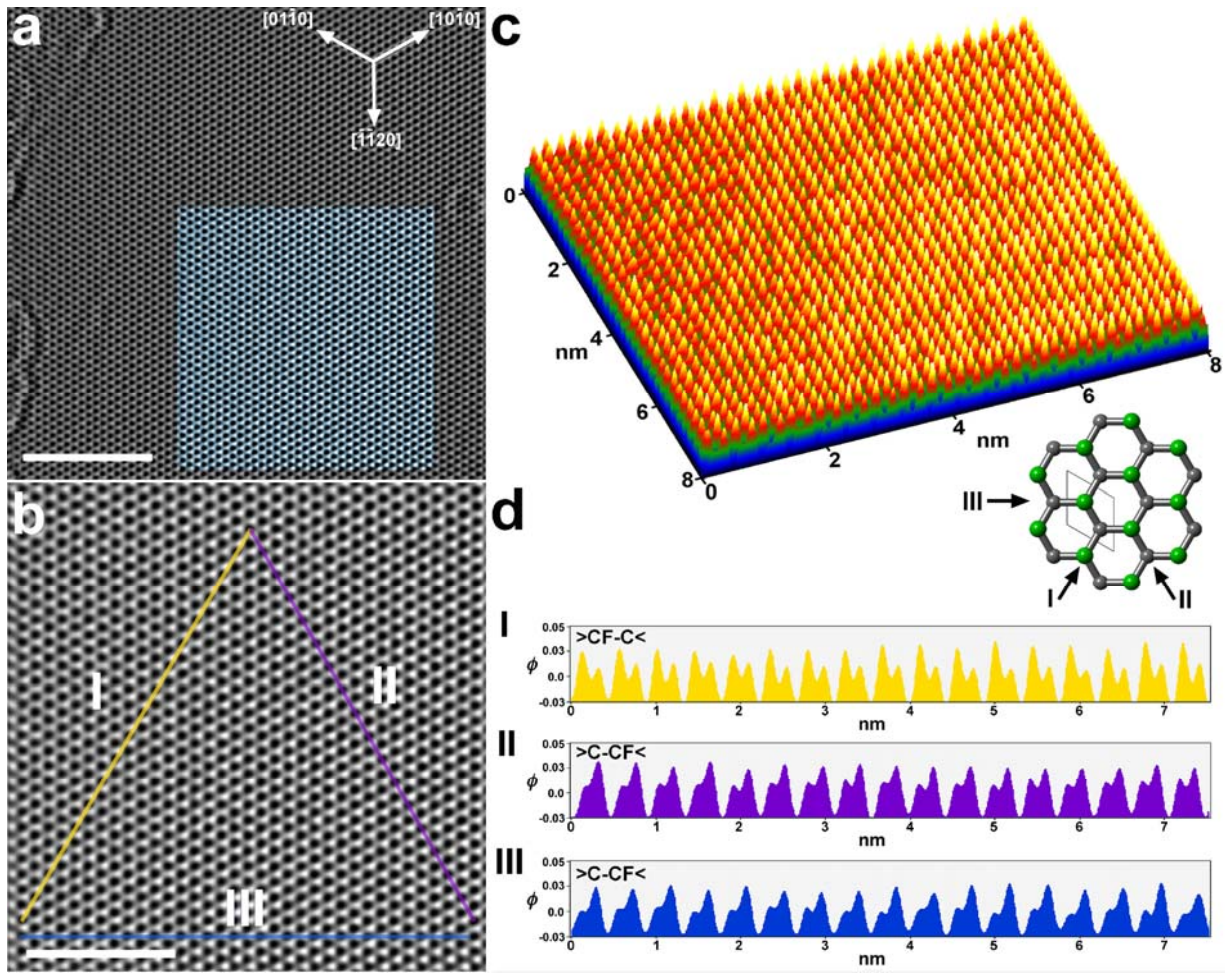

**Figure 4 | Demonstration of long-range order within a 64 nm$^2$ domain of C$_2$F. a,** EWR image of a 250 nm$^2$ sheet of highly ordered sheet C$_2$F with an unrippled 64 nm$^2$ domain highlighted. Outside this domain, ripples are visible (scale bar = 4 nm). **b,** Enlargement of the 64 nm$^2$ domain in **a** exhibiting a high degree of order (scale bar = 2 nm). **c,** Surface plot from **b** in which the orange-yellow apexes correspond to ordered –CF< units within an extended C$_2$F domain with less visible peaks corresponding to single >C- atoms. **d,** Three line profiles (**I-III**) obtained through either >C–CF< or >C–CF< dumb-bells (units of $\Phi$ are rad).

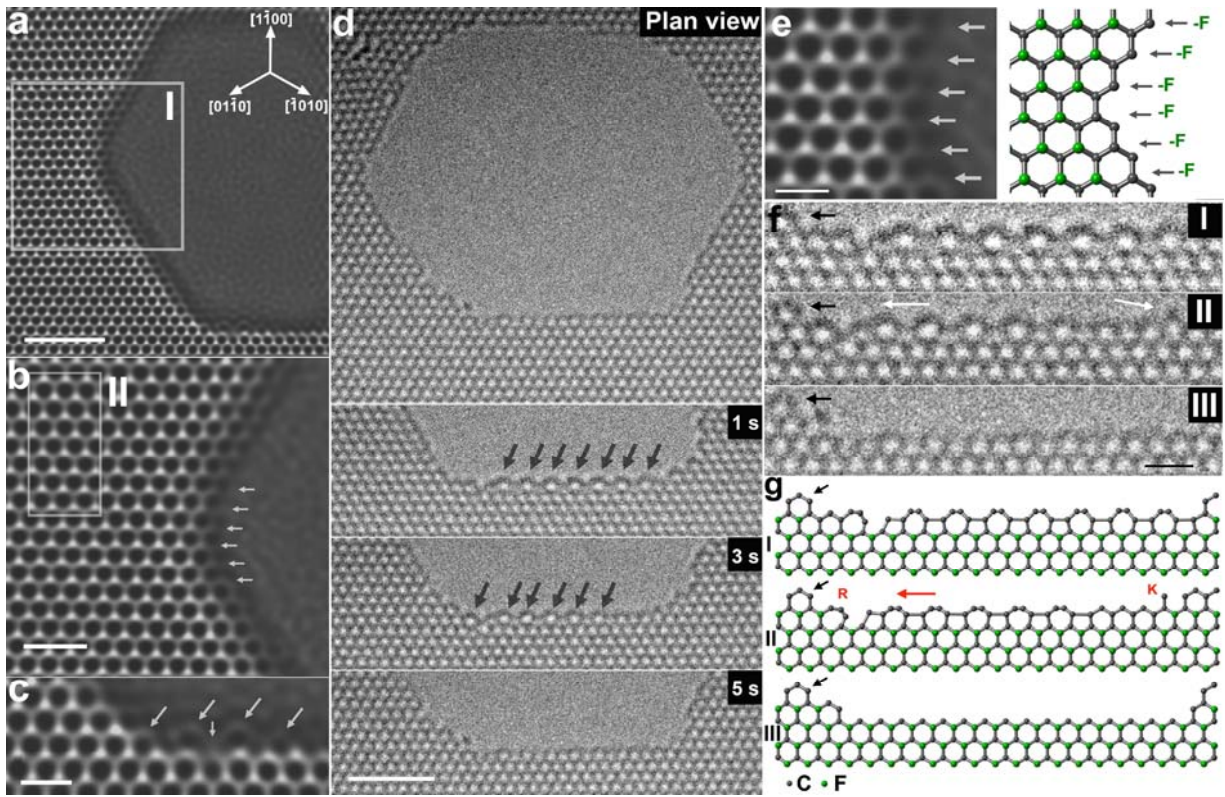

**Fig. 5 | Electron beam modification of a hole formed in $C_2F$ chair a** EWR of a highly ordered domain of $C_2F$ chair during hole formation. On the left, the $C_2F$ chair microstructure is clearly resolved but is more diffuse towards the hole edge (scale bar = 2 nm). **b** Detail from **a** (i.e. region **I**) revealing the microstructure at the hole edge (scale bar = 1 nm). At the edge, the enhanced contrast of the arrowed >C-F positions is reduced due to the progressive elimination of F. **c** Enlarged lower region from **a** (i.e. region **II**) showing diffuse contrast from expanded C-rings at the periphery of the hole (scale bar = 0.5 nm). The arrowed edge features correspond to the dominant microstructure present during focal series acquisition however this blurry region is difficult to interpret. **d** Time-resolved AC-TEM images obtained close to Scherzer imaging conditions. Top image shows a 'plan view' of the entire hole after 5s exposure. Middle panel shows a 1s exposure image with 7-8 expanded C-rings at the edge. At 3s elapsed time five of these rings still remain although two further rings at the extreme left have newly reformed. After 5s total illumination, the entire row is removed leaving an exposed surface layer of $C_2F$ hexagons (scale bar = 2 nm). **e** Detail from **b** (left) and structure model (right) indicating the $C_2F$ chair microstructure during fluorine removal at the edges (scale bar = 0.4 nm). **f** Sequence of unfiltered enlargements from the three bottom insets in d (**I** at 1s; **II**, at 3s and **III** at 5s. Scale bar = 0.5 nm). The C-hexagon indicated by the small black arrow indicates one static point in the image sequence. **g** Three models suggested by the three enlargements **I**-**III** in **f** with the static point indicated. The dominant edge microstructure in **I** and **II** are 7-8 reconstructed zig-zag (or "reczag") units formed from 5- and 7-membered C-rings respectively known to form for carbonaceous graphene although these are probably depleted in F. These units are both mobile and unstable and first rearrange and then are eliminated at **III**. A possible Klein edge is indicated at **K**.

**Supplementary Information**

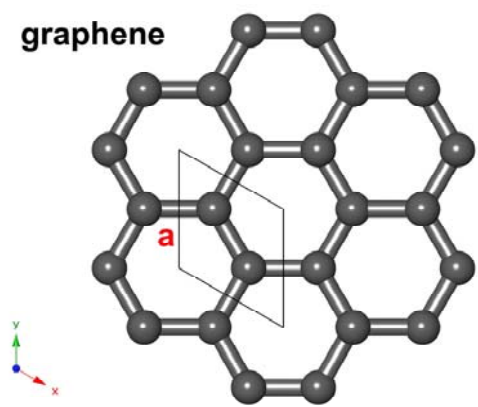
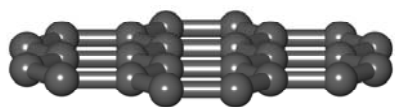

a  graphene
a= 0.246 nm

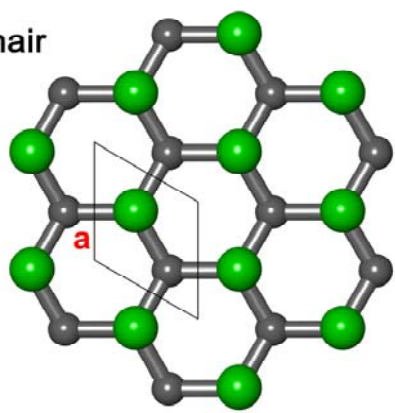
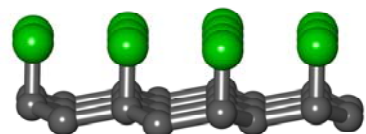

b  C$_2$F chair
a= 0.252 nm

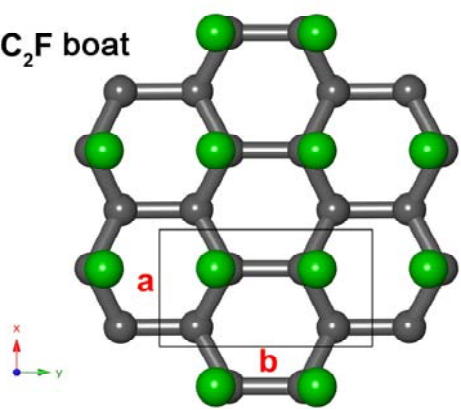
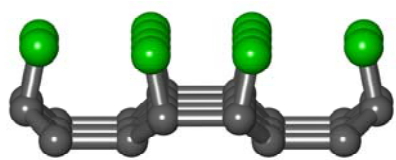

c  C$_2$F boat
a=0.254 nm; b=0.436 nm

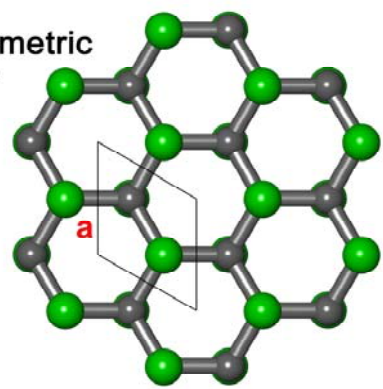
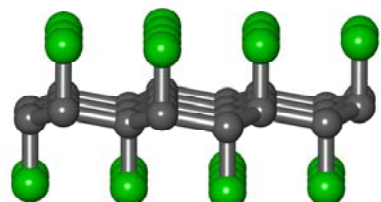

d  Stoichiometric CF
a=0.255 nm

[Cont...]

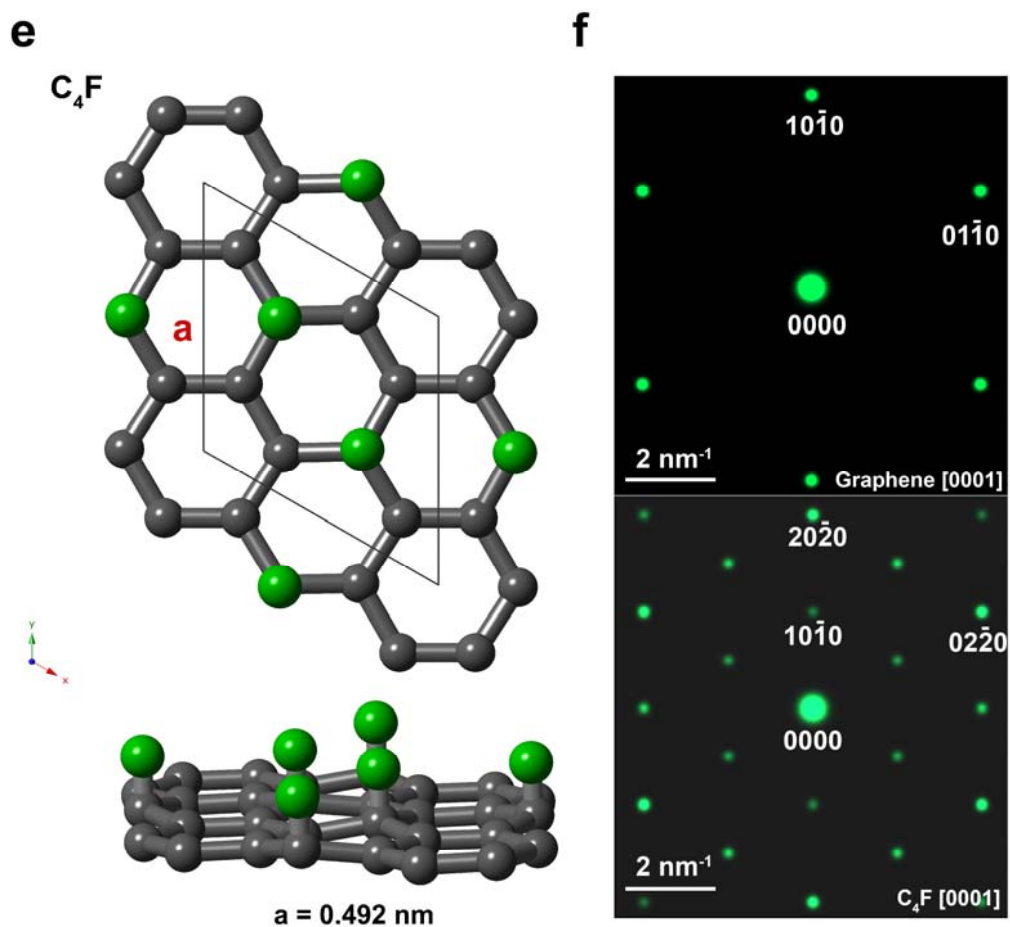

**Supplementary Figure 1 | Complete structure models of graphene, C$_2$F chair, C$_2$F boat, stoichometric fluorographene (CF) and C$_4$F a** Plan-view (i.e. [001] projection) and side-on perspective views of carbonaceous graphene. **b** Plan view and side-on perspective view of mono-sided C$_2$F chair. **c** Plan view and side-on perspective view of mono-sided C$_2$F boat. **d** Plan view and side-on perspective view of mono-sided stoichiometric fluorographene (CF). **e** Plan view and side-on perspective view of C$_4$F with a double lattice parameter (i.e. with respect to unfluorinated graphene). **f** Simulated electron diffraction patterns of unfluorinated graphene (top) and C$_4$F (bottom). Only the latter 2D phase has a supercell formed as a result of alternating fluorination although this is not observed in C$_2$F chair. All unit cells are as defined by Şahin (Ref. 10, main Communication).

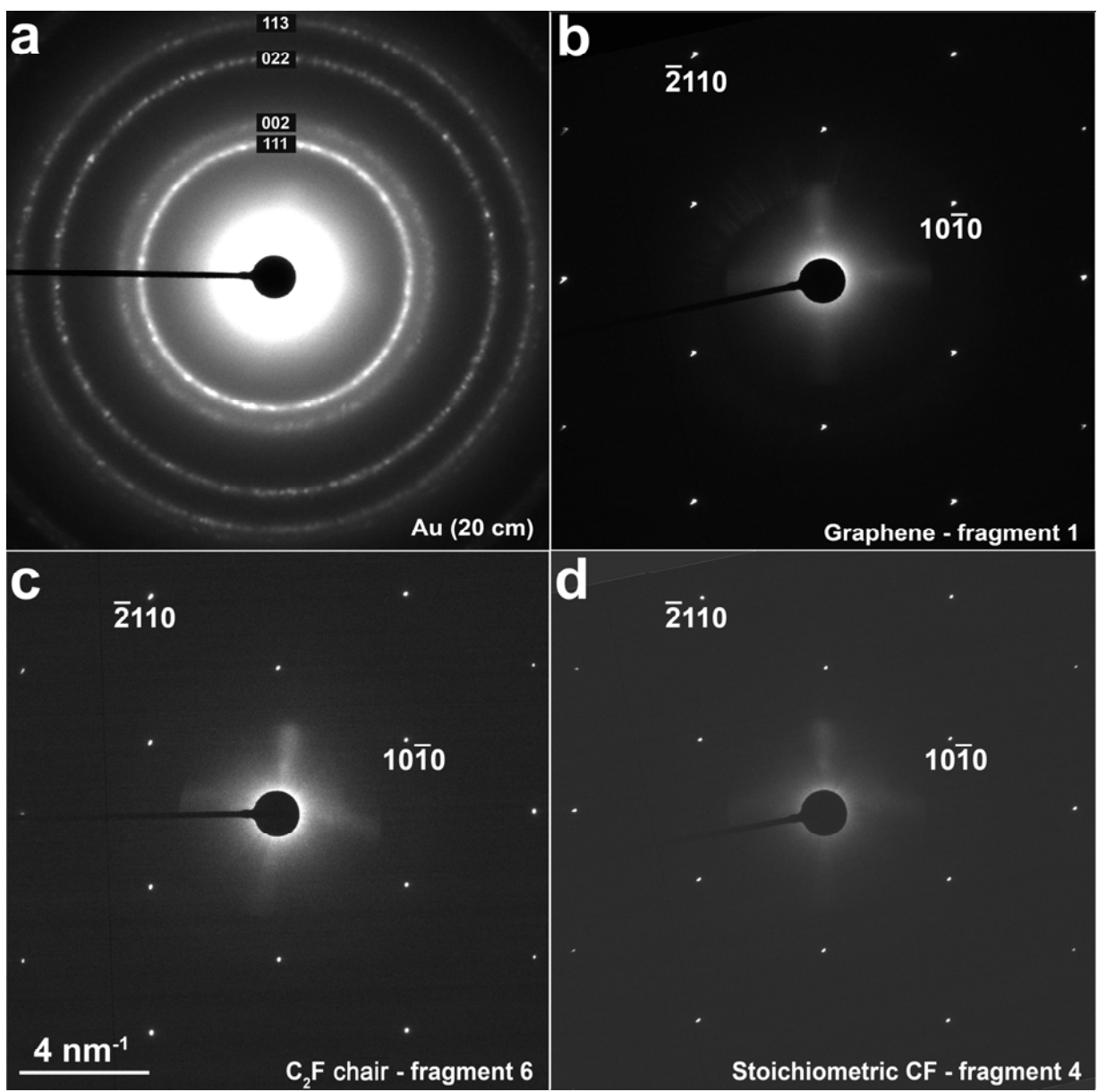

**Supplementary Figure 2 | Electron diffraction patterns from polycrystalline gold, graphene, *chair*-C$_2$F and stoichiometric CF a** Electron diffraction ring pattern obtained from polycrystalline gold sample on carbon. The most prominent rings are $d_{111}$ (0.2355 nm); $d_{002}$ (0.2040 nm); $d_{022}$ (0.1442 nm); $d_{113}$ (0.1230 nm) (source: calculated for $Fm\bar{3}m$ Au collection code 44362-ICSD (http://icsd.cds.rsc.org)). These were used to calibrate the camera length of the ARM200F to within at least 0.6% precision. **b** Electron diffraction pattern from [0001] projection native graphene from Fragment 1 (see also Supplementary Table 2); **c** Electron diffraction pattern from [0001] projection C$_2$F chair from Fragment 6 (see also Supplementary Table 2); **d** Electron diffraction pattern from [0001] projection stoichiometric CF from Fragment 4 (see also Supplementary Table 2).

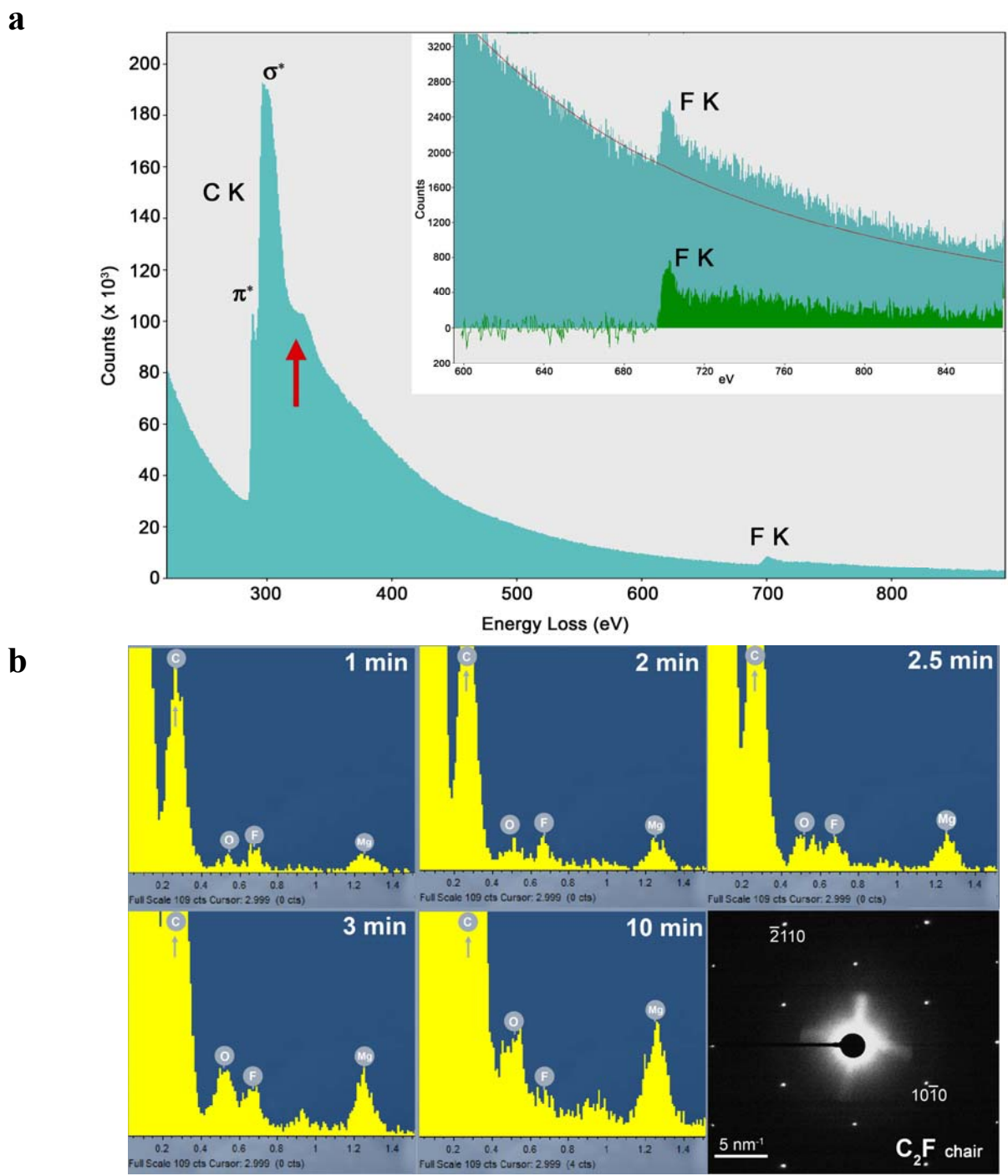

**Supplementary Figure 3 | EELS and EDX of fluorinated CVD graphene membrane prepared on a TEM grid a** EELS spectrum obtained from first fluorinated graphene sample several weeks after EWR image analysis. This spectrum was obtained from a clump of carbonaceous material and still shows significant evidence of fluorination with a ratio of F to C calculated at 10% however this may not be representative of the local F:C ratio due to the observed rapid contamination of the sample during EELS acquisition. It was noted that the $\sigma^*$ component of the C K edge (arrowed) increased markedly during spectrum acquisition suggesting rapid specimen contamination. **b** Representative time-resolved EDX spectra obtained from a single $C_2F$ chair fragment (i.e. Fragment 5, Supplementary Table 2) with corresponding inset ED pattern. The buildup of C impurities and other impurities (i.e. Mg, O) is also evident during extended spectrum acquisition. All similar $C_2F$ chair fragments) presented with similar spectra whereas unfluorinated graphene fragments (see also Supplementary Table 2) showed no evidence of fluorination.

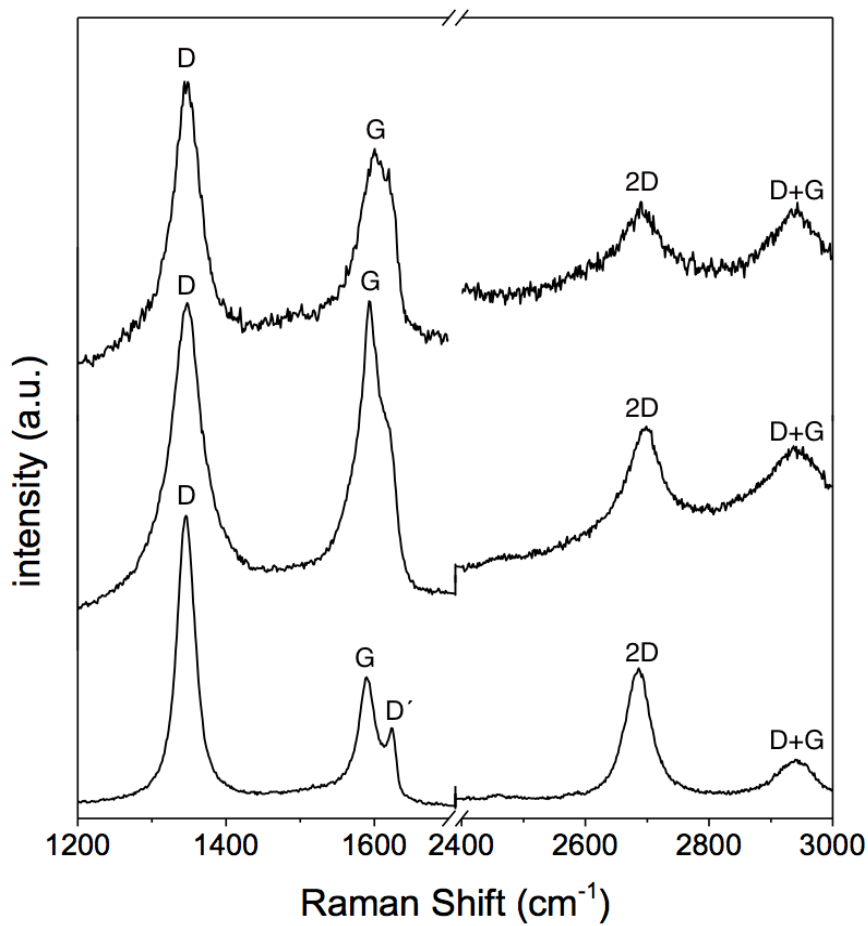

**Supplementary Figure 4 | Raman analysis of fluorinated CVD graphene membrane prepared on a TEM grid** This figure shows the Raman spectrum of a partially fluorinated graphene from three different regions of the sample. Unlike fully fluorinated graphene the Raman spectrum of partially fluorinated samples are highly non-uniform and different regions of the samples shows different D-peak to G-peak intensity ratio (and also different 2D and D+G modes). This evidence also supports the existence of several types of variably fluorinated domains in partially fluorinated samples (see Supplementary Tables 1 and 2).

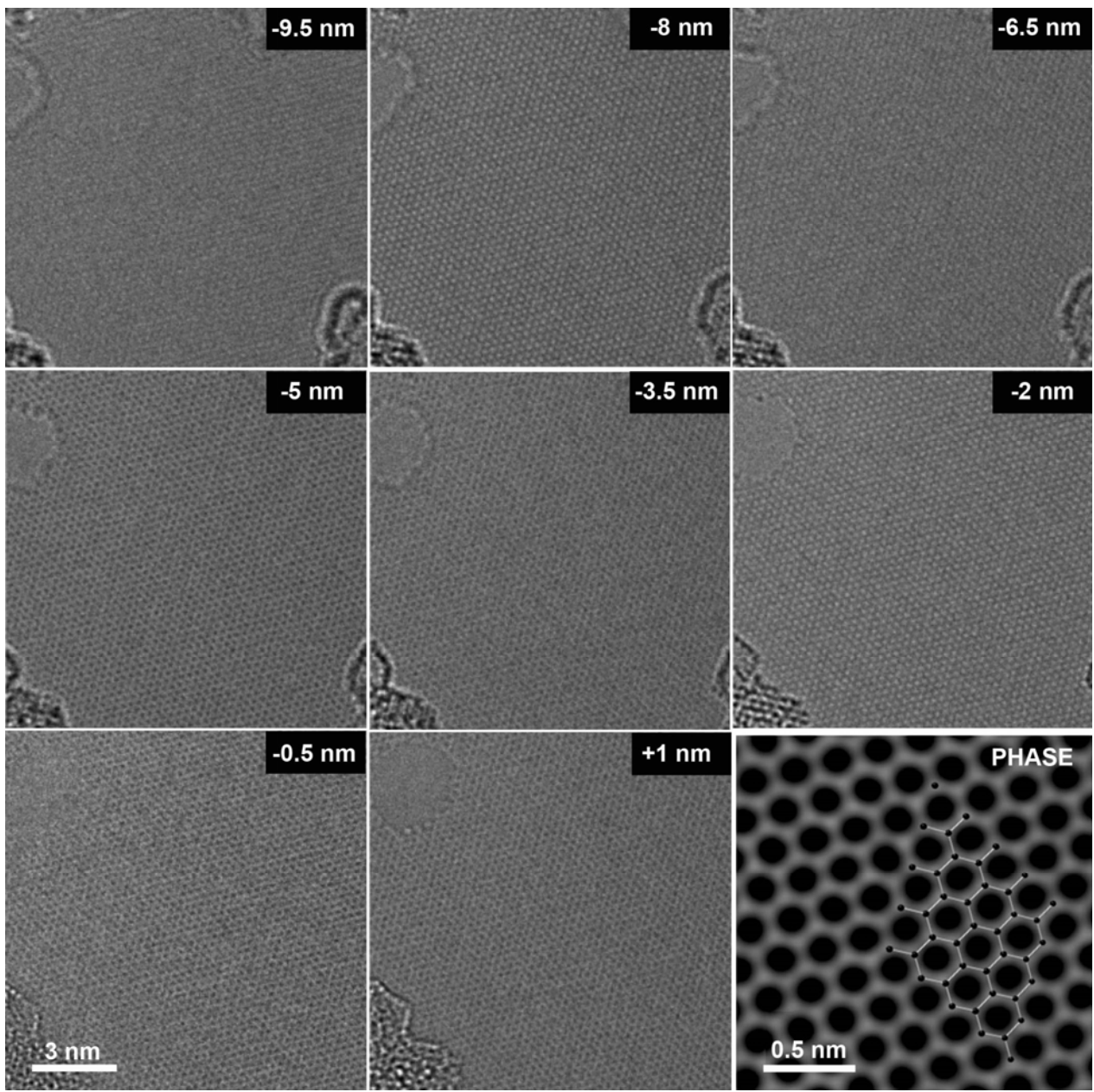

**Supplementary Figure 5 | Focal series and detail of an Exit-Wave Reconstruction (EWR) phase image for pristine graphene** Eight images extracted from a 34-member focal series obtained from a mostly pristine domain of graphene. The EWR phase image (bottom right) was reconstructed for a defect-free region in roughly the centre of the -2 nm defocus image.

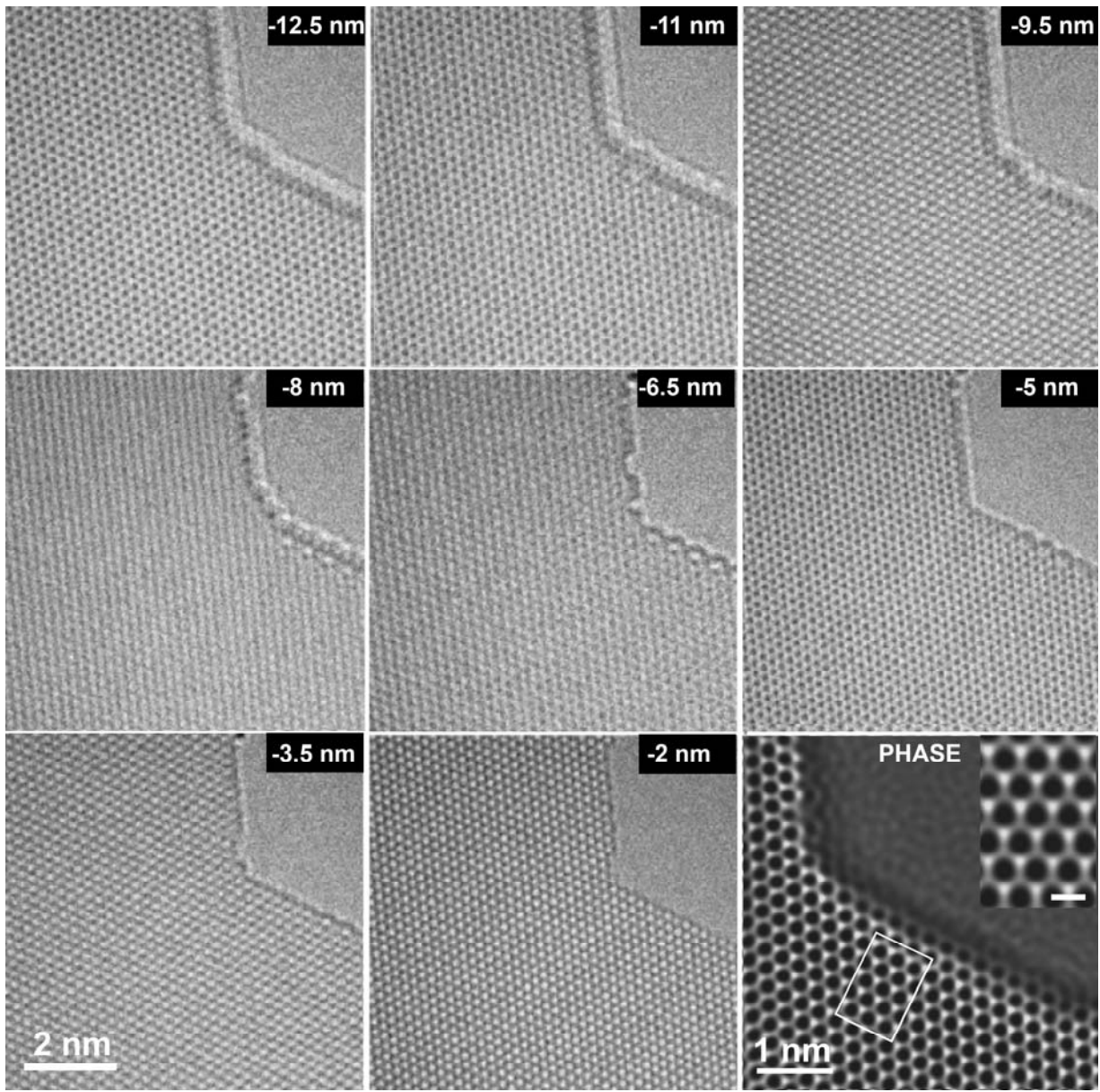

**Supplementary Figure 6 | Focal series and detail of an Exit-Wave Reconstruction (EWR) phase image for $C_2F$ chair.** Eight images extracted from a 34-member focal series obtained from a region of highly ordered $C_2F$ chair containing a hole. The EWR phase image (bottom right) was reconstructed for a defect-free region towards the edge of the hole. The detail inset into the bottom right phase image (scale bar = 0.25 nm) reveals the high degree of order about ~0.4 nm from the edge.

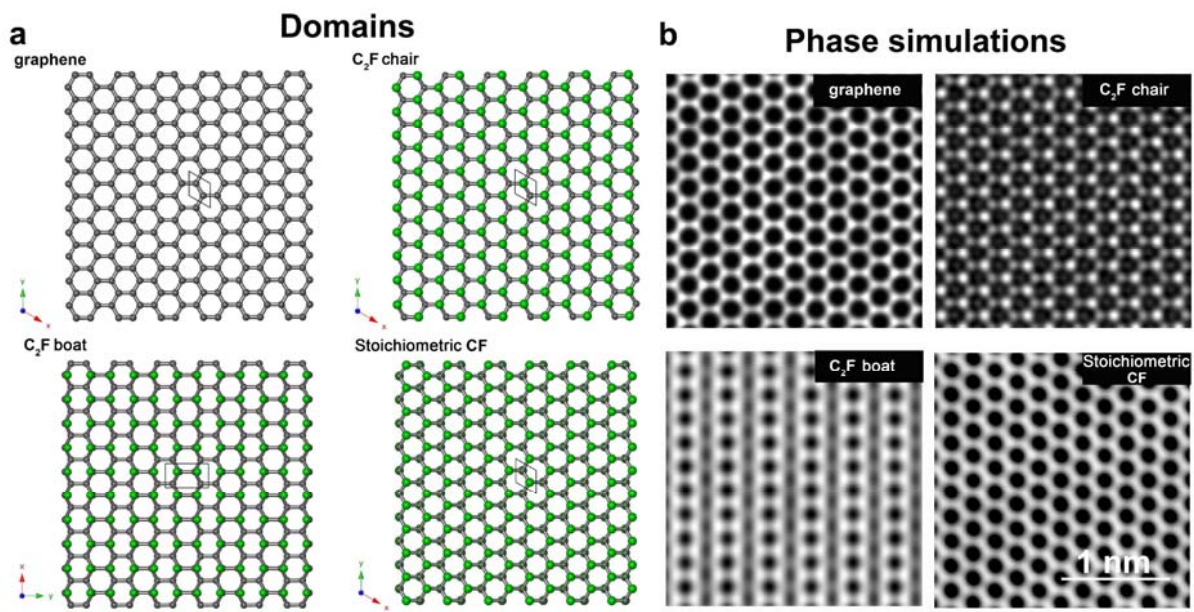

**Supplementary Figure 7 | Comparison of EWR phase image simulations produced for *chair*-$C_2F$ chair, $C_2F$ boat, and Stoichometric fluorographene (CF) a** Structure models of four domains of pristine graphene (top left), $C_2F$ chair (top right), $C_2F$ boat (bottom left) and stoichometric fluorographene (bottom right). **b** Corresponding phase image simulations of pristine graphene (top left), $C_2F$ chair (top right), $C_2F$ boat (bottom left) and stoichometric fluorographene (bottom right) performed using a fast multslice algorithm (see Methods section, main Communication, for more details) and the domain models in **a**. The EWR phase image contrast for $C_2F$ boat clearly cannot be confused with that of either $C_2F$ chair or stoichiometric CF.

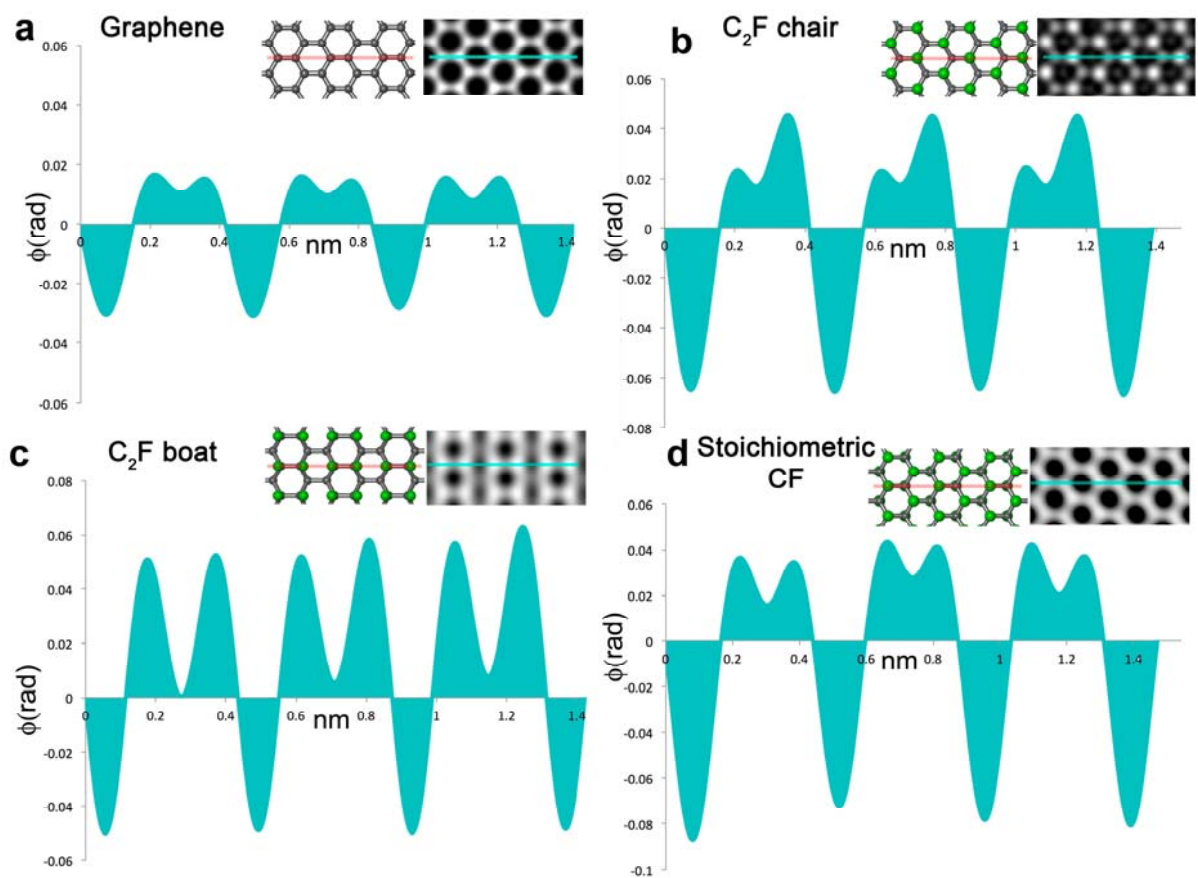

**Supplementary Figure 8 | Comparison of EWR phase image simulations line profiles produced for $C_2F$ chair, $C_2F$ boat, and stoichometric fluorographene (CF) a** Using a small domain extracted from the corresponding EWR phase simulation (i.e. Supplementary Figure 4, **a** and **b**) line profiles through three C-C dumb-bells were obtained. Taking into account the +ve and −ve components of the phase shift a net phase shift for the individual C atoms is ~0.05 rad. **b** as for **a** but this time the corresponding 'saw-tooth' phase shifts through a single C atom and a >C-F pair for three >C-CF< $C_2F$ chair dumb-bells is plotted. Note that the −ve component of the phase shift is convoluted with that for both the single C atom and the >C-F pair. The maximum net phase shift of ~0.1 rad is that expected for a >C-F pair (cf. **c** and **d**). **c** As for **a** and **b** but this time line profiles through three >CF-CF< dumb-bells (in which the F atoms are on one side only) for $C_2F$ boat are plotted. Each column has a net phase shift of ~0.1 rad and there is no 'saw-tooth' contrast as for $C_2F$ chair. **d** As for **a-c** but this time three line profiles through three >CF-CF< dumb-bells (in which F atoms are on alternating sides) for stoichiometric. While the contrast of stoichiometric CF is superficially similar to the contrast observed for pristine graphene, the net phase shift of ~0.1 rad per >CF- par is double that calculated for graphene (*cf.* **a**).

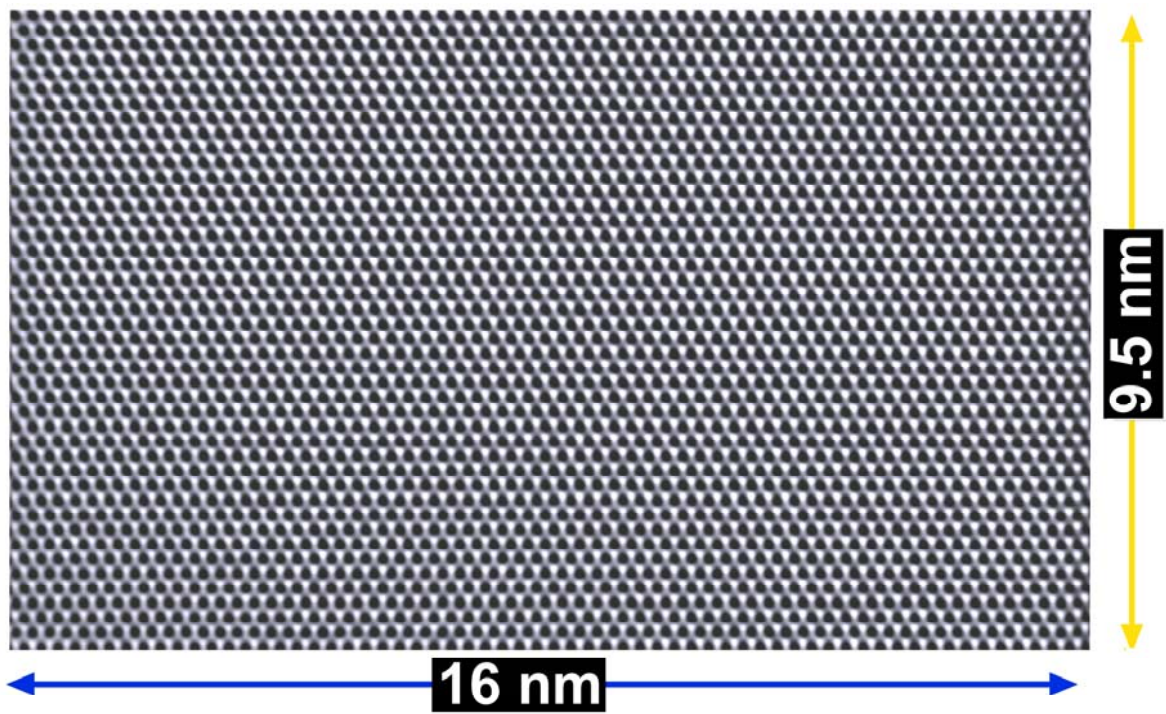

**Supplementary Figure 9 | Demonstration of extended long-range order within a ~150 nm$^2$ domain of $C_2F$ chair** Completely pristine 16 × 9.5 nm domain of ordered $C_2F$ chair imaged by EWR. Diffuse rippling is just visible on the left hand side of this image but this is not topological in nature. Viewing this image at a glancing angle in any lattice direction reveals no evidence topological defects or associated evidence of strain.

**Supplementary Table 1 | Compiled lattice parameters of graphene and DFT optimised and experimental fluorinated graphene derivatives** DFT-refined lattice parameters for graphene, $C_2F$ chair, $C_2F$ boat, stoichiometric CF respectively.[S1] Also shown, for comparison, is the experimental reported lattice parameters for stoichiometric.[S2]

| Phase | Lattice Parameter | (nm) | % Deviation from Graphene | % Deviation from $C_2F$ Chair |
|---|---|---|---|---|
| Graphene | a | 0.246 | – | -2.38 |
| $C_2F$ Chair | a | 0.252 | 2.44 | – |
| $C_2F$ Boat | a | 0.254 | – | – |
| $C_2F$ Boat | b | 0.436 | – | – |
| Stoichiometric CF (DFT) | a | 0.255 | 3.66 | 1.19 |
| Stoichiometric CF (Exp) | a | 0.248 | 0.81 | -1.59 |

**Supplementary Table 2 | Lattice parameter analysis of second fluorinated graphene sample**
Schematic figure illustrating the comparative distribution of graphene (**Gr**, inset graphic) versus stoichiometric CF (**Stoich. CF**, inset graphic) and $C_2F$ chair (**Ch-$C_2F$**, inset graphic) in 18 fragments studied from the second fluorinated sample. Fragment No., averaged $d_{100}$ (i.e. $10\bar{1}0$ d-spacing), identified phase, derived $a$ parameter and % deviation of $a$ from the expected values listed in Supplementary Table 1. The estimated precision of the Au-calibrated lattice parameter measurements from electron diffraction (e.g. Supplementary Figs. 2a-d) is at least 0.6%.

| Fragment No. | Average $d_{(100)}$ (nm) | Identified Phase | $a$ (nm) | % Deviation from identified phase |
|---|---|---|---|---|
| 1 | 0.21342 | Graphene | 0.24643 | 0.18 |
| 2 | 0.21389 | Graphene | 0.24697 | 0.40 |
| 3 | 0.22184 | Stoichiometric CF | 0.25616 | 0.45 |
| 4 | 0.22220 | Stoichiometric CF | 0.25657 | 0.62 |
| 5 | 0.21672 | $C_2F$ Chair | 0.25025 | -0.69 |
| 6 | 0.21757 | $C_2F$ Chair | 0.25122 | -0.31 |
| 7 | 0.21854 | $C_2F$ Chair | 0.25235 | 0.14 |
| 8 | 0.20940 | Graphene | 0.24179 | -1.71 |
| 9 | 0.20956 | Graphene | 0.24198 | -1.63 |
| 10 | 0.21383 | Graphene | 0.24691 | 0.37 |
| 11 | 0.21690 | $C_2F$ Chair | 0.25046 | 0.61 |
| 12 | 0.21797 | $C_2F$ Chair | 0.25169 | 0.12 |
| 13 | 0.21461 | Graphene | 0.24781 | 0.74 |
| 14 | 0.21559 | Graphene | 0.24894 | 1.20 |
| 15 | 0.21786 | $C_2F$ Chair | 0.25157 | -0.17 |
| 16 | 0.21687 | $C_2F$ Chair | 0.25041 | -0.63 |
| 17 | 0.21793 | $C_2F$ Chair | 0.25164 | -0.14 |
| 18 | 0.21764 | $C_2F$ Chair | 0.25131 | -0.27 |

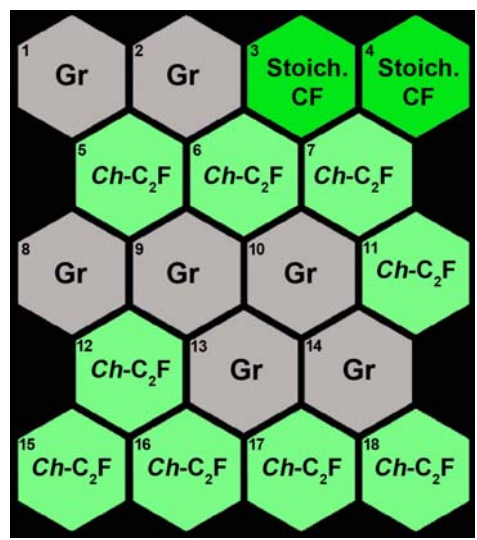